\documentclass[12pt,twocolumn,reprint,floatfix]{revtex4-1}
\usepackage{amsmath,latexsym}    
\usepackage{graphicx}   
\usepackage{float}
\usepackage{hyperref}   
\usepackage{epstopdf}
\begin{document}
\linespread{1.3} 
\title{Suppression of the superconductivity in ultrathin amorphous Mo$_{78}$Ge$_{22}$ thin films observed by STM}
\author{D. Lotnyk}
\author{O. Onufriienko}
\author{T. Samuely}
\author{O. Shylenko}
\author{V. Komanick\'{y}}
\author{P. Szabo}
\author{A. Feher}
\author{P. Samuely}
\affiliation{Centre of Ultra Low Temperature Physics, Institute of
Experimental Physics, Slovak Academy of Sciences, and P. J.
Safarik University, SK-04001, Kosice, Slovakia}

\date{\today}

\begin{abstract}
In contact with a superconductor a normal metal modifies its
properties due to Andreev reflection. In the current work the
local density of states (LDOS) of superconductor - normal metal
Mo$_{78}$Ge$_{22}$ - Au bilayers are studied by means of STM
applied from the Au side. Three bilayers have been prepared on
silicate glass substrate consisting of 100, 10 and 5~nm MoGe thin
films covered always by 5 nm Au layer. The tunneling spectra were
measured at temperatures from 0.5~K to 7~K. The two-dimensional
cross-correlation between topography and normalized zero-bias
conductance (ZBC) indicates a proximity effect between 100 and
10~nm MoGe thin films and Au layer where a superconducting gap
slightly smaller than that of bulk MoGe is observed. The effect of
the thinnest 5~nm MoGe layer on Au leads to much smaller gap
moreover the LDOS reveals almost completely suppressed coherence
peaks. This is attributed to a strong pair-breaking effect of
spin-flip processes at the interface between MoGe films and the
substrate.
\end{abstract}
\maketitle

\section{Introduction}
Disorder and reduced dimensionality affect the physical properties
of metallic systems in several ways~\cite{Anderson79,
Altshuler80}. Strong diffusion leads to localization of electrons
and a related enhancement of the Coulomb interaction. In a system
of lower dimensions, the coupling to disorder increases and
pronounced effects are expected.

Amorphous MoGe thin films are suitable system for studying the
interplay between superconductivity and reduced dimensionality.
Several transport experiments showed a reduction of the
superconducting transition temperature $T_c$~\cite{Graybeal,
Strongin} which could lead to superconductor - insulator
transition~\cite{Haviland, Szabo16}.

In a contact with a normal metal the superconductor wave function
is changed by the proximity effect~\cite{DeGennes}.
On the normal-metal side a minigap or suppressed superconducting gap is
induced~\cite{Belzig}.

In this work, the local density of states (LDOS) has been measured
by means of the scanning-tunneling-microscope (STM) technique. The
two-dimensional cross correlation between the topography of
locally varying Au thickness and the tunneling zero-bias
conductance (ZBC) for 10~nm and 100~nm MoGe bilayer samples
indicate a presence of the proximity effect in the LDOS of Au
overlayer. The LDOS spectra on Au reveal the gap which is only
slightly suppressed in comparison with that in bulk MoGe. It means
that down to 10 nm thickness the superconducting properties of
MoGe are not changing. In the bilayer with 5~nm MoGe the tunneling
spectrum shows a high ZBC and strongly suppressed coherence peaks.
Such a strong suppression of superconducting LDOS in the same 5 nm
Au layer cannot come just from proximity effect. Also there is no
correlation between the topography and ZBC. Then, the proximity
effect is not sufficient to explain the spectra measured on Au but
superconductivity in the 5~nm MoGe is strongly suppressed. We
consider a strong pair breaking due to the spin-flip processes at
the interface between MoGe and the glass substrate. Consequently,
superconductivity in MoGe will be supressed as well by an inverse
proximity effect.

\section{Experimental details \label{ss:expdet}}
\begin{figure}[H]
 \begin{minipage}{0.95\columnwidth}
    \raggedright{
        \includegraphics[width=0.87\columnwidth]{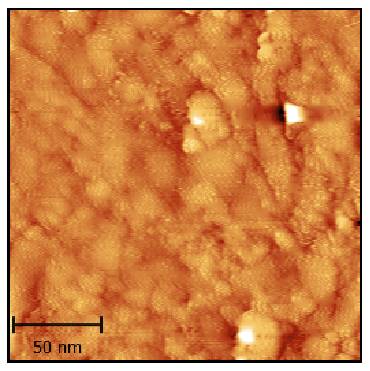}
         }
    \end{minipage}
\\ \vskip\baselineskip
   \begin{minipage}{0.95\columnwidth}
    \begin{center}
        \includegraphics[width=\columnwidth]{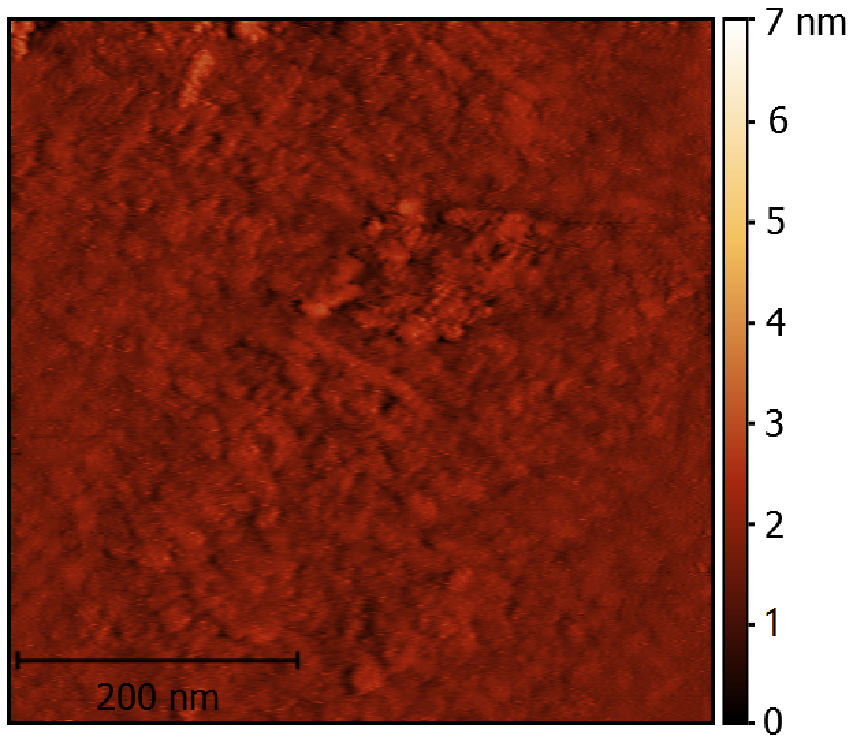}
        \end{center}
    \end{minipage}
\\ \vskip\baselineskip
    \begin{minipage}{0.95\columnwidth}
    \raggedright{
        \includegraphics[width=0.87\columnwidth]{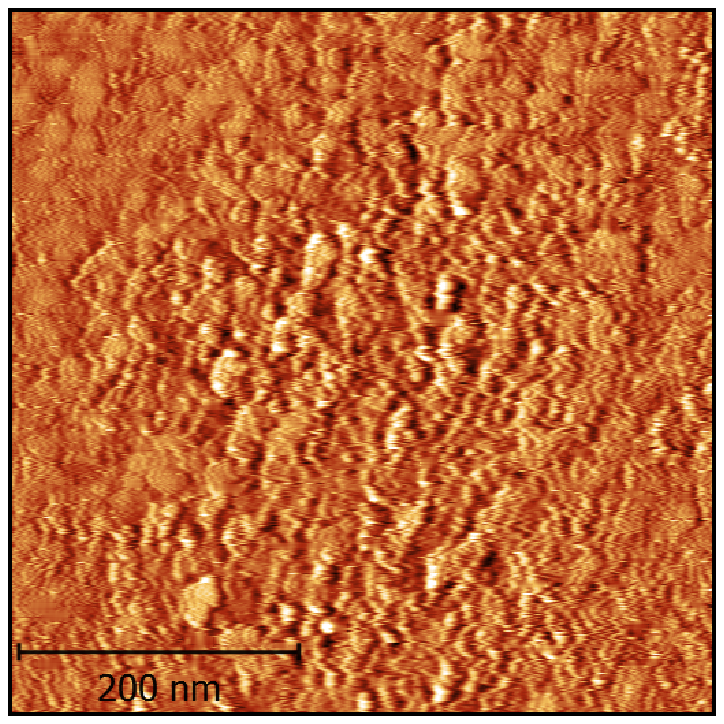}
    }
    \end{minipage}
\caption{Topographies acquired by STM of (top) 5~nm thick MoGe
(250$\times$250~nm$^2$, 10~mV, 0.53~K); (middle) 10~nm
(500$\times$500~nm$^2$, 10~mV, 0.53~K); and (bottom) 100~nm
(500$\times$500~nm$^2$, 10~mV, 0.53~K).}\label{fig:topo}
\end{figure}

A magnetron sputtering technique was used to produce MoGe thin
films with thicknesses of 5, 10, and 100~nm covered by Au layer
with thickness of 5~nm. The films were deposited on commercially
available silicate glass slides of 10~mm x 10~mm x 0.5~mm size.
MoGe was sputtered on the rotating sample holder from a commercial
single composite target with a 99.9+~\% purity. The sputtering
conditions for MoGe were: chamber pressure
3.7$\times$10$^{-10}$~Pa, argon pressure 2.2$\times$10$^{-5}$~Pa,
DC target power 100~W, substrate temperature 22~$^{\circ}$C, the
sputtering rate is~0.59~\AA s$^{-1}$. Immediately after MoGe
deposition, the Au thin film was sputtered with the rate~4.54~\AA
s$^{-1}$ to prevent the oxidation. The thickness of the sputtered
films was controlled by quartz crystal microbalance device. The
topography of the samples is shown in Fig.~\ref{fig:topo}. RMS
roughnesses of 5~nm, 10~nm and 100~nm MoGe thin films are 1.09~nm,
0.38~nm and 0.76~nm, respectively. The sizes of topography are
500~nm x 500~nm for 10 and 100~nm thin films and 250~nm x 250~nm
for 5~nm film.

The scanning tunneling microscopy and spectroscopy experiments
were performed by means of a homemade STM head~\cite{tsam2012}
controlled by the Nanotec's Dulcinea SPM

electronics. The head is inserted in a commercial Janis SSV
cryomagnetic system with $^3$He refrigerator and the Au tip was
prepared in-situ by repetitive impaling into the bulk Au sample
and subsequent slow retraction.

The procedure was repeated until the current – position dependence
exhibited the conductance quantization and single-atom contact
phenomena inherent for gold~\cite{Agrait}. The sample was grounded
while bias voltage was applied to the tip with the tunneling
resistance 10~M$\Omega$.

The individual spectra have been measured in temperature range
between 0.5~K and 7~K. At the lowest temperature we have also
performed the current imaging tunneling spectroscopy (CITS)
providing the spectral maps at each point of the topography.

\section{Results and discussion}
The experimental conductance spectra were analyzed assuming normal
metal - insulator - superconductor (NIS) contact described by
eq.~\cite{Tinkham}:

\begin{equation}
\centering
 \frac{dI_{nis}}{dV}=G_{nn}\int_{-\infty}^{\infty}\frac{N_s(E)}{N_n(0)}\left[-\frac{\partial f(E+eV)}{\partial (eV)}\right], \label{eq:difcond}
\end{equation}

where $e$ is a charge of electron, $I_{nis}$ is a tunneling
current, $V$ is a bias voltage between Au tip and the sample,
$dI/dV$ is experimental differential conductance, $G_{nn}$ --
normal-state differential conductance, $N_s(E)$ is superconducting
density of states (SDOS), $N_n(0)$ is the normal-state density of
states at Fermi level, and $f$ is the Fermi-Dirac distribution
function. The superconducting density of states was taken in the
Dynes form \cite{Dynes}:
\begin{equation}
\centering
 N_s(eV)=N_n(0)\mathrm{Re}\left[\frac{eV+i\Gamma}{\sqrt{(eV+i\Gamma)^2-\Delta^2}}\right], \label{eq:dynes}
\end{equation}
where $\Delta$ is a superconducting gap and $\Gamma$ is the
broadening parameter.

Each curve was normalized to the normal state spectrum. Since the
Au tip possess a constant density of states ($N_n(0)=const$),
consequently, each of differential conductance spectra reflects
the SDOS of the MoGe, smeared by 2$k_BT$ in energy at the
measuring temperature. In the low temperature limit (at
$k_BT\ll\Delta$), the differential conductance corresponds
directly the SDOS.

\begin{figure}[h]
 \begin{minipage}{0.95\columnwidth}
    \begin{center}
        \includegraphics[width=0.9\columnwidth]{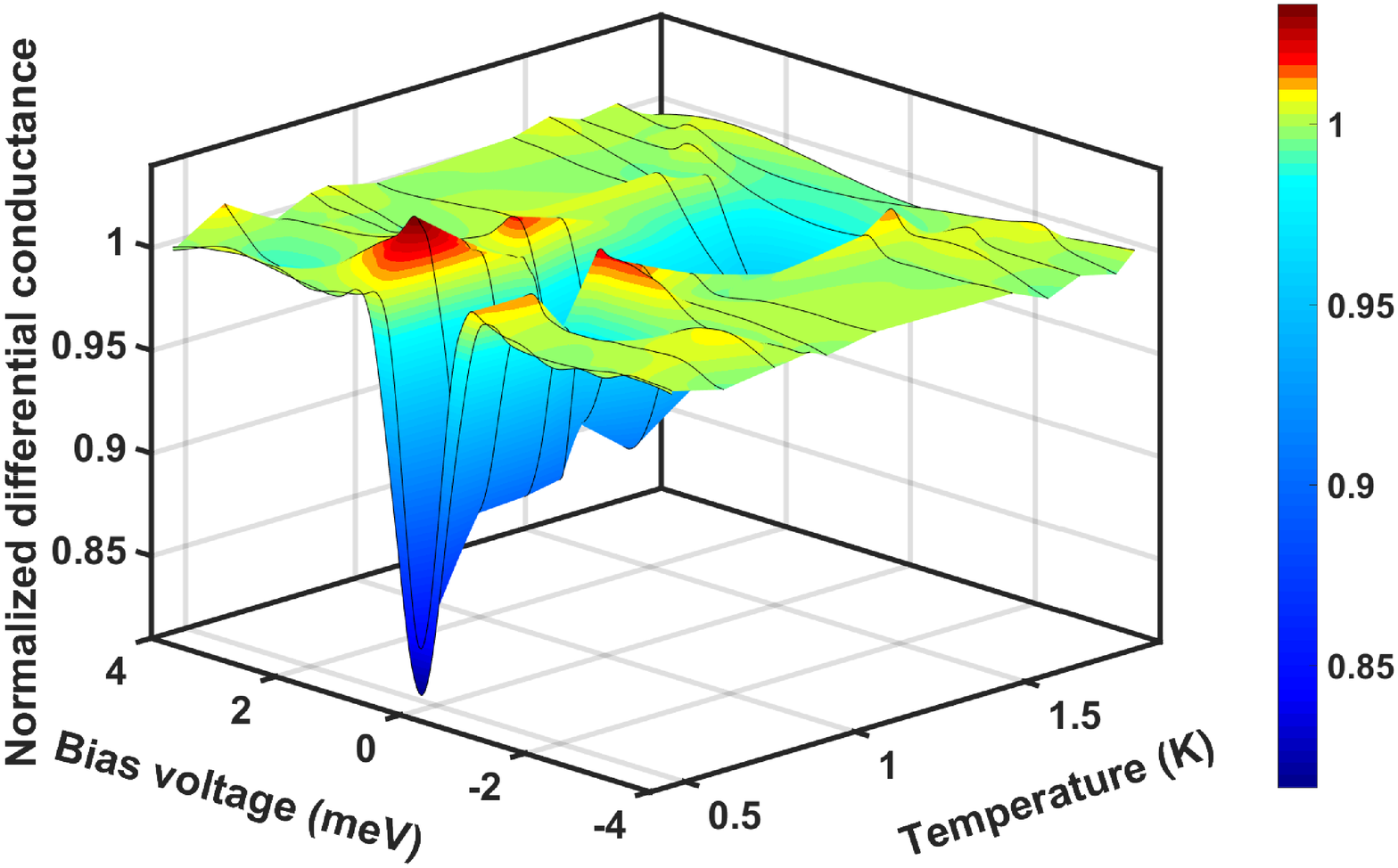}
        \end{center}
    \end{minipage}
   \begin{minipage}{0.95\columnwidth}
    \begin{center}
        \includegraphics[width=0.9\columnwidth]{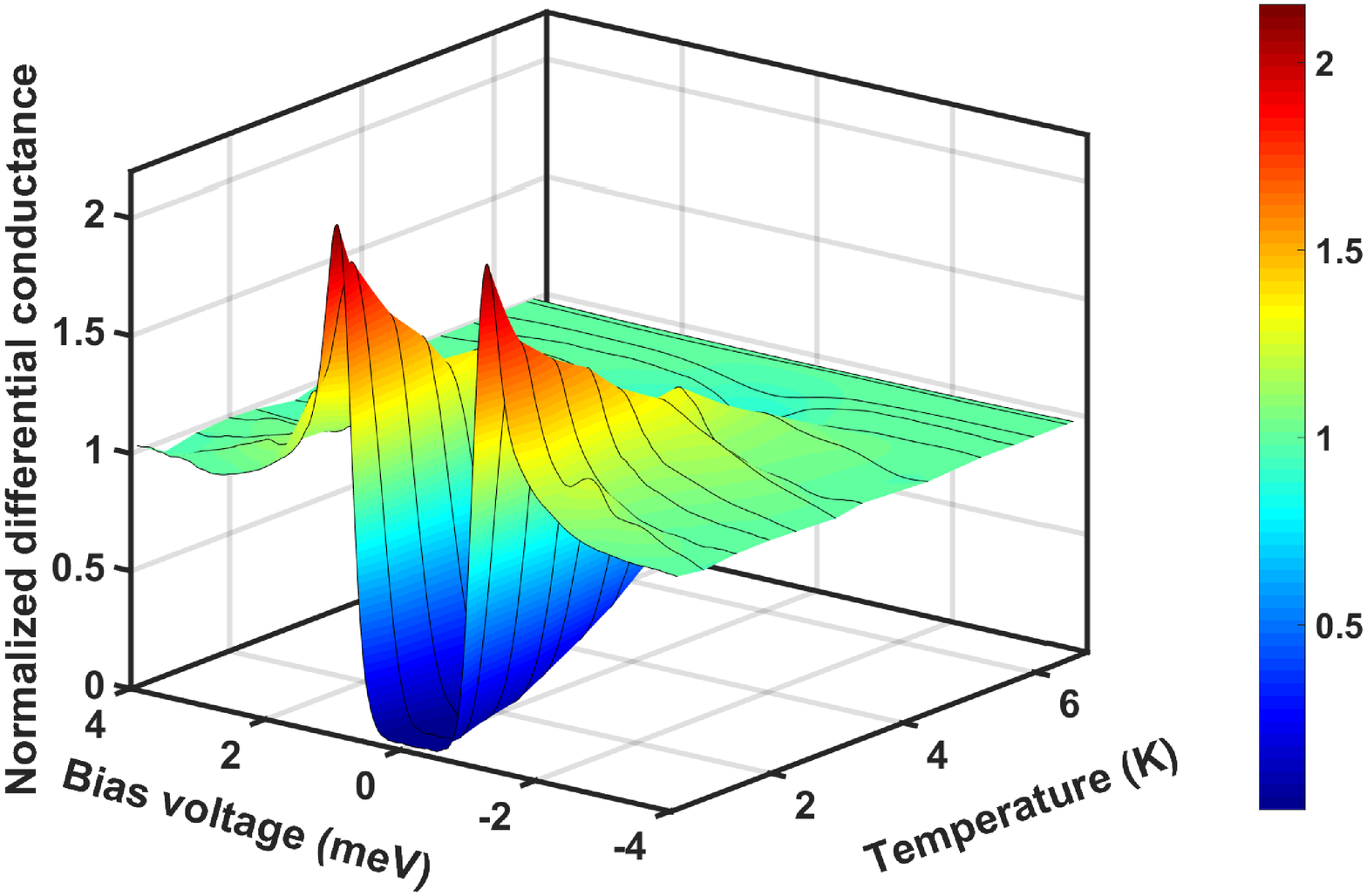}
        \end{center}
    \end{minipage}
    \begin{minipage}{0.95\columnwidth}
    \begin{center}
        \includegraphics[width=0.9\columnwidth]{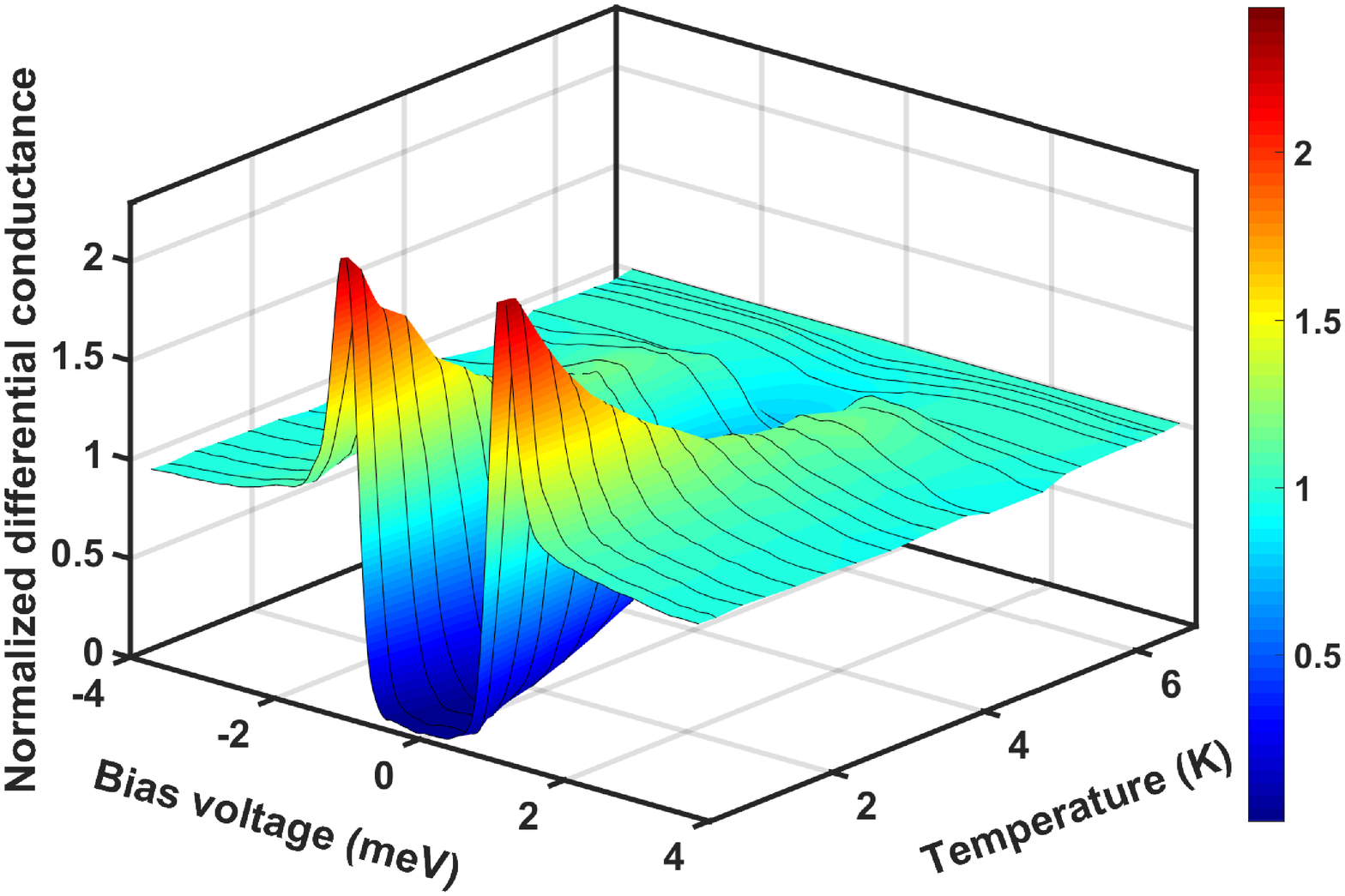}
        \end{center}
    \end{minipage}
\caption{Normalized differential conductance spectra acquired by
STM spectroscopy measured between the Au tip and the MoGe samples
5~nm (top), 10~nm (middle), and 100~nm (bottom) in zero magnetic
field at different temperatures between 0.5~K and
6.5~K.}\label{fig:didv}
\end{figure}

In Figure~\ref{fig:didv}, the three-dimensional plots of the
normalized tunneling conductance spectra for the MoGe samples
versus temperature are shown. Each series of conductance spectra
was reproduced in, at least, two different points of topography
(Fig.~\ref{fig:topo}).

\begin{figure}[t]
\includegraphics[width=0.95\columnwidth]{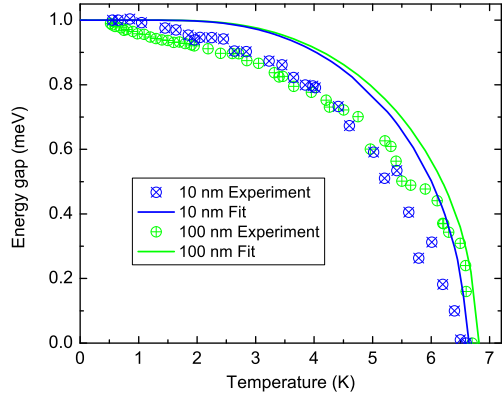}
 \caption{ \label{fig_delta} Temperature dependencies of the superconducting energy gap
 for 10~nm (green) and 100~nm (blue) MoGe thin films obtained using Dynes approximation.
 Dots denote experimental values, lines denote fit according to BCS theory~\eqref{eq:temp_delta}.}
\end{figure}

All curves were fitted according to the eq.~\eqref{eq:difcond} in
order to obtain values of $\Delta$ and $\Gamma$. For the samples
with MoGe thicknesses of 10~nm and 100~nm the results are plotted
in Fig.~\ref{fig_delta}. The broadening parameter $\Gamma$ does
not exceed 0.1~meV. The results for 10~nm and 100~nm samples
reveal a similar behavior with the critical temperatures 6.7~K,
6.6~K and superconducting gaps 1.02~meV, 1.05~meV, respectively.
Due to the heavily smeared spectra of the 5~nm MoGe sample, it was
impossible to fit them by eq.~\ref{eq:difcond}. Directly fom the
Fig.~\ref{fig:didv} we estimated critical temperature for 5~nm
MoGe as $T_c$~=1.9~K, where the ZBC is close to unity. The value
of $\Delta(0)$~=~0.7~meV as a distance between coherence peaks was
estimated as an upper limit. The superconducting coupling ratio
$2\Delta(0)/k_BT_c$ values equal to 3.5 for the both 100 and 10~nm
samples. This is very close to the values obtained on similar MoGe
films by infrared spectroscopy \cite{Tashiro}. In contrast, the
5~nm sample reveals a very high coupling ratio
$2\Delta(0)/k_BT_c$~=~8.

The temperature dependence of $\Delta(T)$ was compared with
theoretical curve (lines in Fig.~\ref{fig_delta}) according to the
BCS model \cite{Abrikosov}:

\begin{equation}
\centering
\ln\frac{\Delta(0)}{\Delta(T)}=\int_{0}^{\infty}\frac{1-\mathrm{tanh}[(\zeta^2+\Delta^2)^{1/2}/2T]}{(\zeta^2+\Delta^2)^{1/2}}\mathrm{d\zeta},
\label{eq:temp_delta}
\end{equation}

where $\Delta(0)$ is the superconducting gap at zero temperature.
One can see, that experimental data (dots) deviates from the
theoretical prediction (lines). We suggest that such behavior
could be explained by the influence of proximity effect.

Since the measurements were performed on the normal side of the
metal-superconductor bilayer, the Andreev
reflection~\cite{Andreev} would modify the local density of
states. Instead of the true superconducting gap, the so-called
minigap $\Delta_n$ \cite{Belzig} would be obtained from the
experiment. Due to difficulty of numerical analysis \cite{Belzig},
in the current paper we perform only qualitative analysis of the
experimental results. We can define two different coherence
length. The first one $\xi_s$ is for superconducting layer and the
second $\xi_n$ for adjacent normal Au overlayer:

\begin{equation}
\centering
 \xi_{n,s}=\sqrt{\frac{\hbar D_{n,s}}{2\Delta_{n,s}}},
 \label{eq:xsi}
\end{equation}

where $\xi_{n,s}$ are the normal and superconducting coherence
lengths, $D=\frac{1}{3}v_Fl_{el}$ is the diffusive constant with
$v_F$ is Fermi velocity, and $l_{el}$ is the elastic mean free
path. In ref.~\cite{Gupta} the normal coherence length on a
similar bilayer Nb/Au has been estimated as $\xi_n$=60~nm. The
ratio of minigap to the real superconducting gap $\Delta_n/\Delta$
depends on $\xi_n$ and the thickness of the normal layer
$L_n$~\cite{Belzig}:
\onecolumngrid 

\begin{figure}[H]
\begin{minipage}{0.5\columnwidth}
    \begin{center}
        \includegraphics[width=\columnwidth]{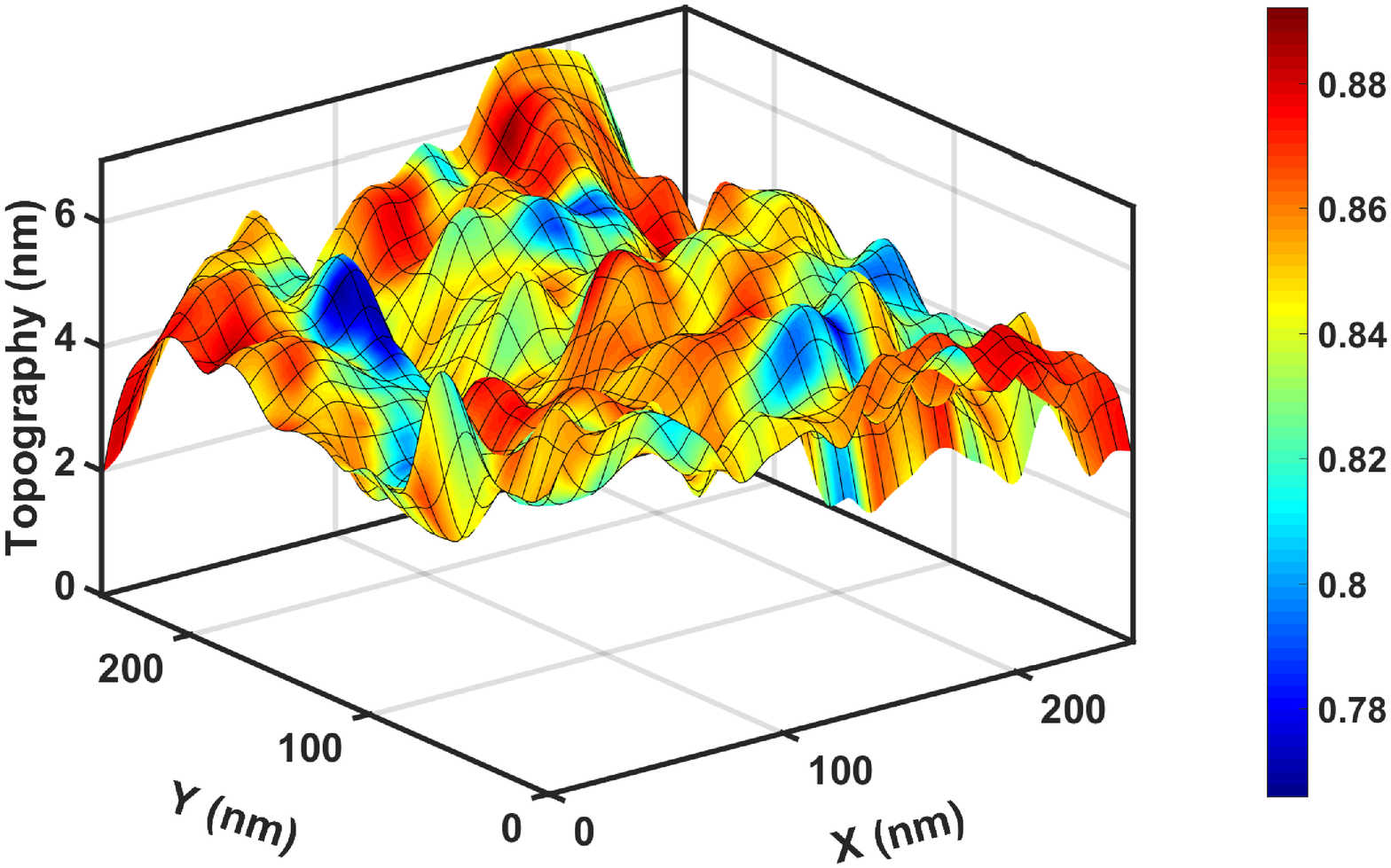}
        (a)
        \end{center}
    \end{minipage}
    \qquad
    \begin{minipage}{0.45\columnwidth}
    \begin{center}
        \includegraphics[width=0.95\columnwidth]{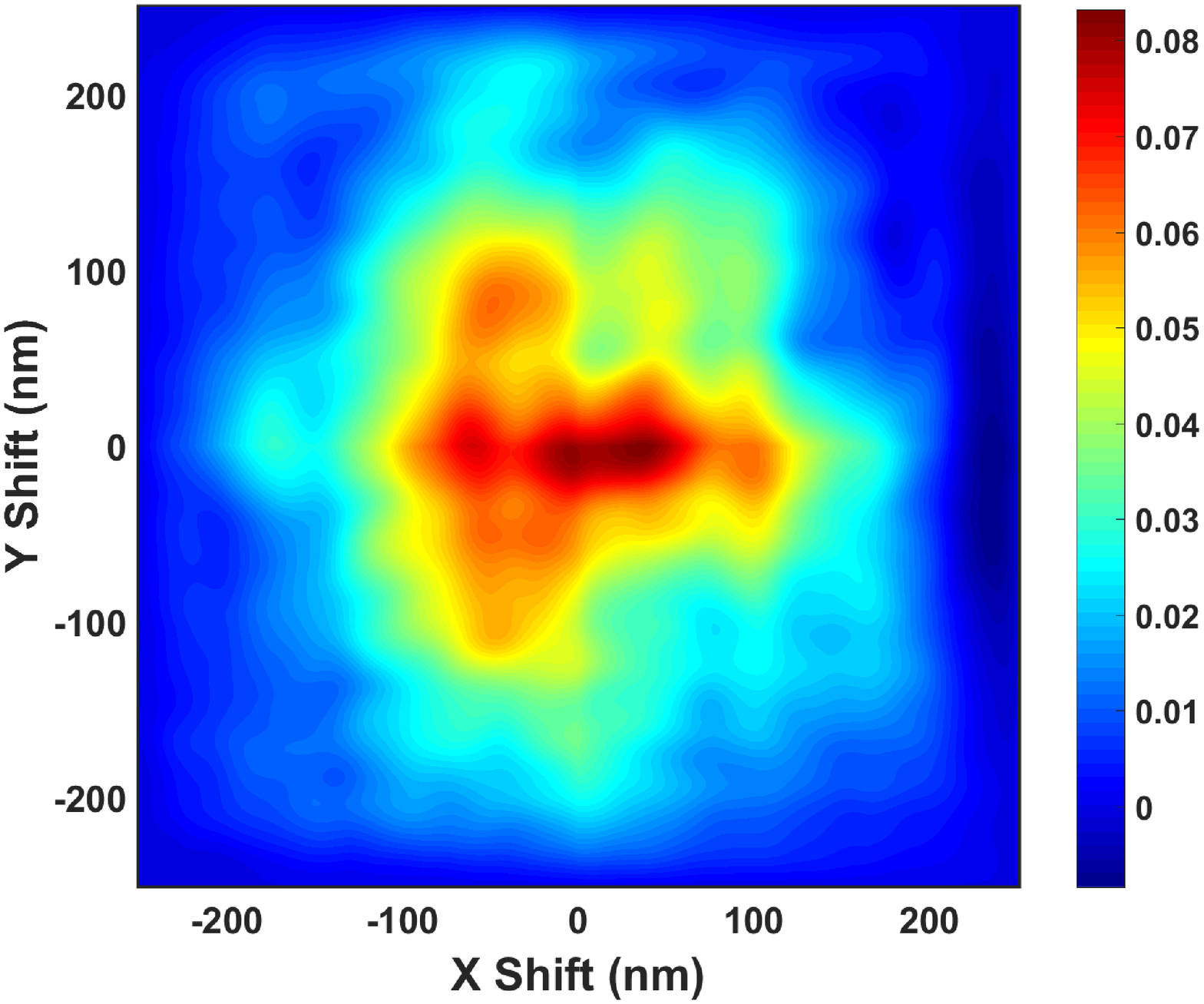}
        (b)
        \end{center}
    \end{minipage}
\\ \vskip\baselineskip
\begin{minipage}{0.46\columnwidth}
    \begin{center}
        \includegraphics[width=\columnwidth]{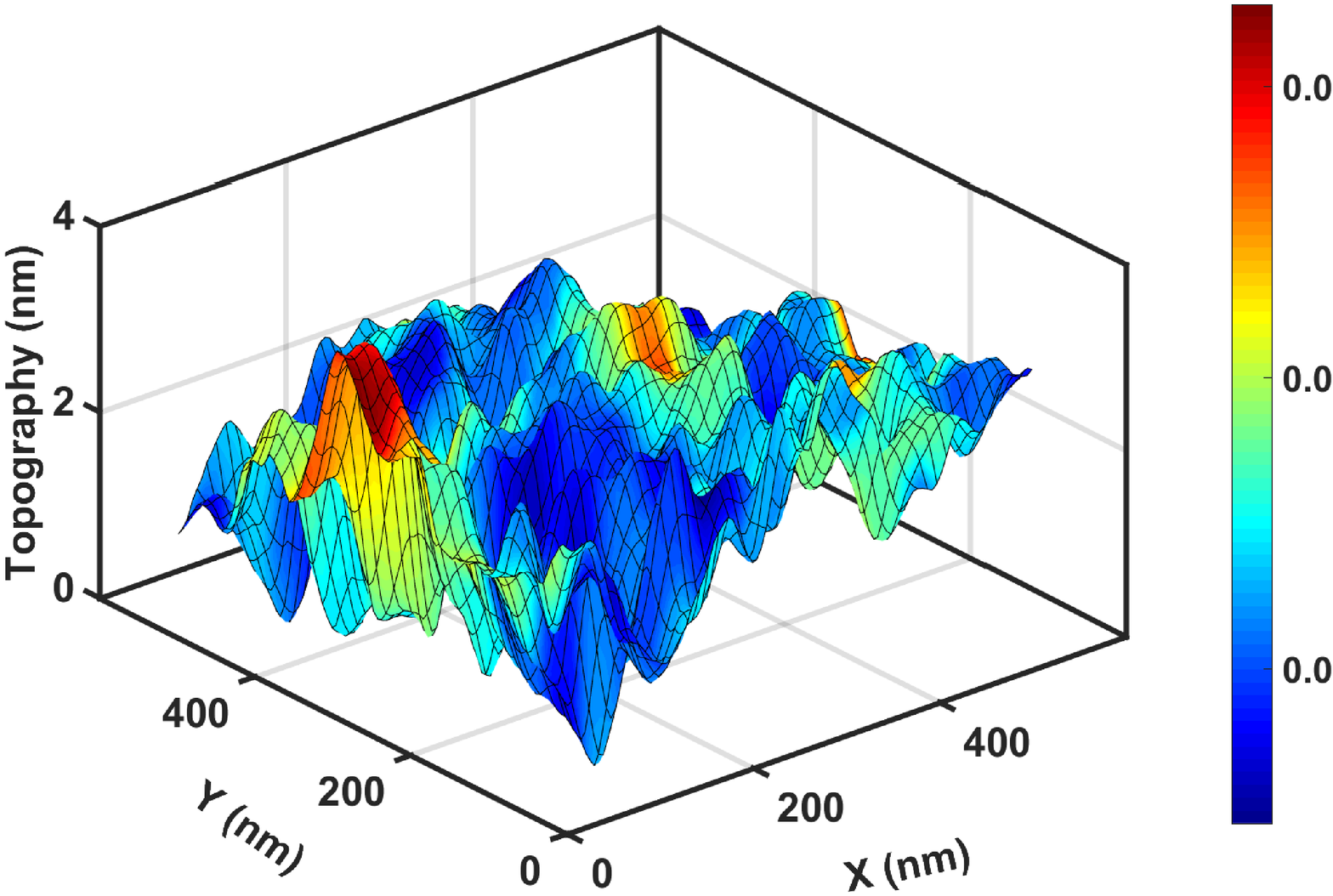}
        (c)
        \end{center}
    \end{minipage}
    \qquad\hskip 8mm plus 5mm
   \begin{minipage}{0.45\columnwidth}
    \begin{center}
        \includegraphics[width=0.95\columnwidth]{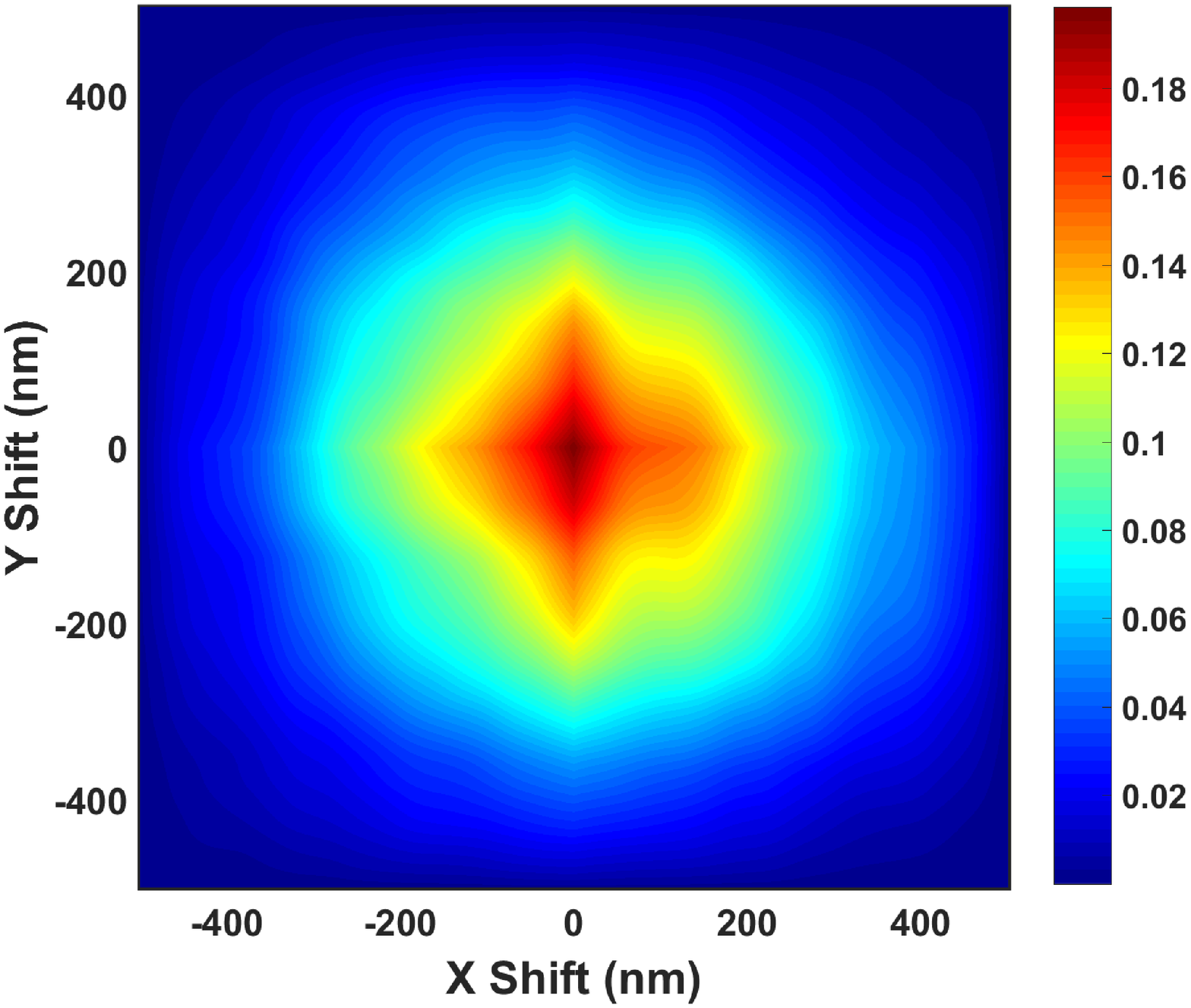}
        (d)
        \end{center}
    \end{minipage}
\\ \vskip\baselineskip
\begin{minipage}{0.5\columnwidth}
    \begin{center}
        \includegraphics[width=0.95\columnwidth]{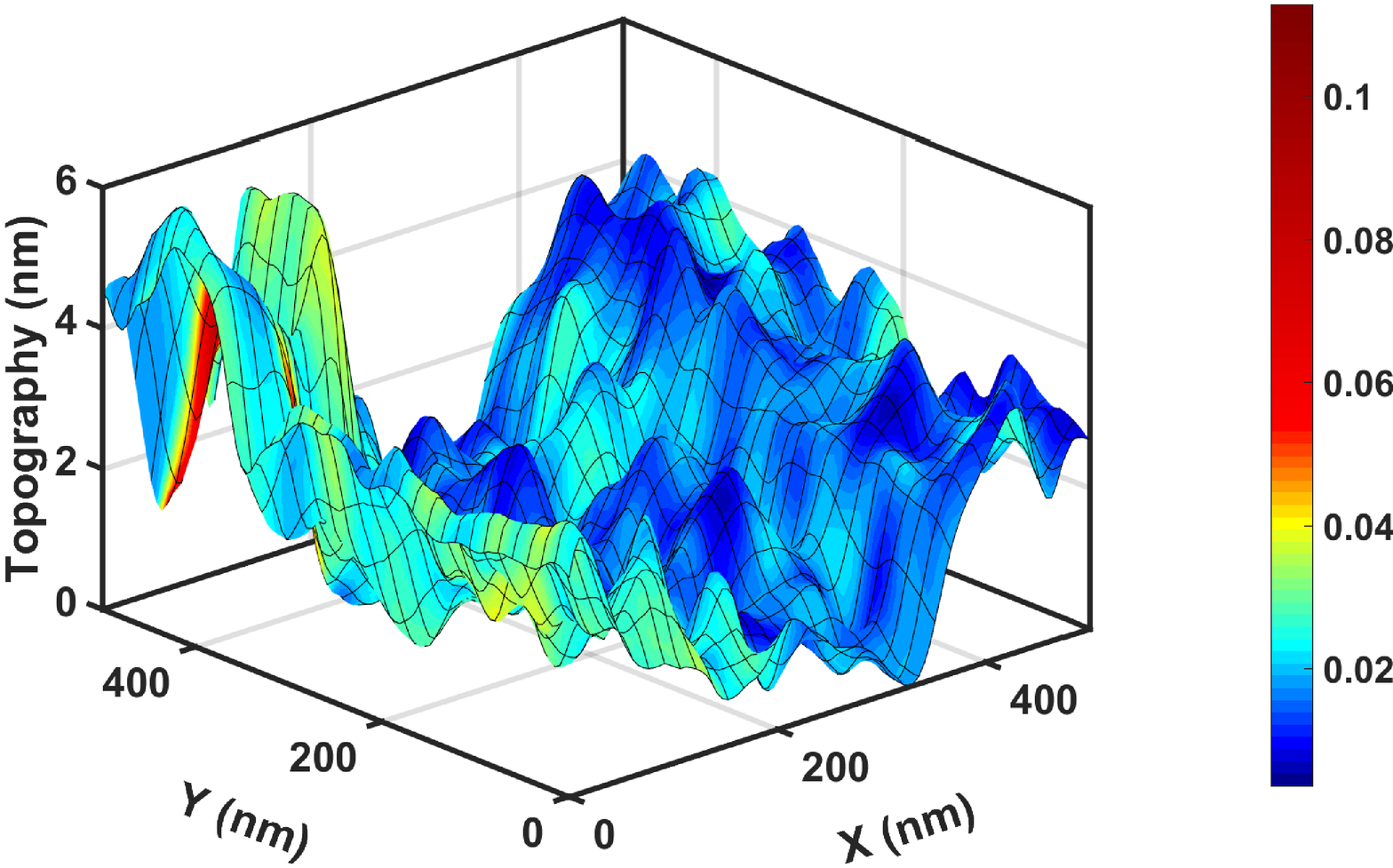}
        (e)
        \end{center}
    \end{minipage}
    \qquad
    \begin{minipage}{0.45\columnwidth}
    \begin{center}
        \includegraphics[width=0.95\columnwidth]{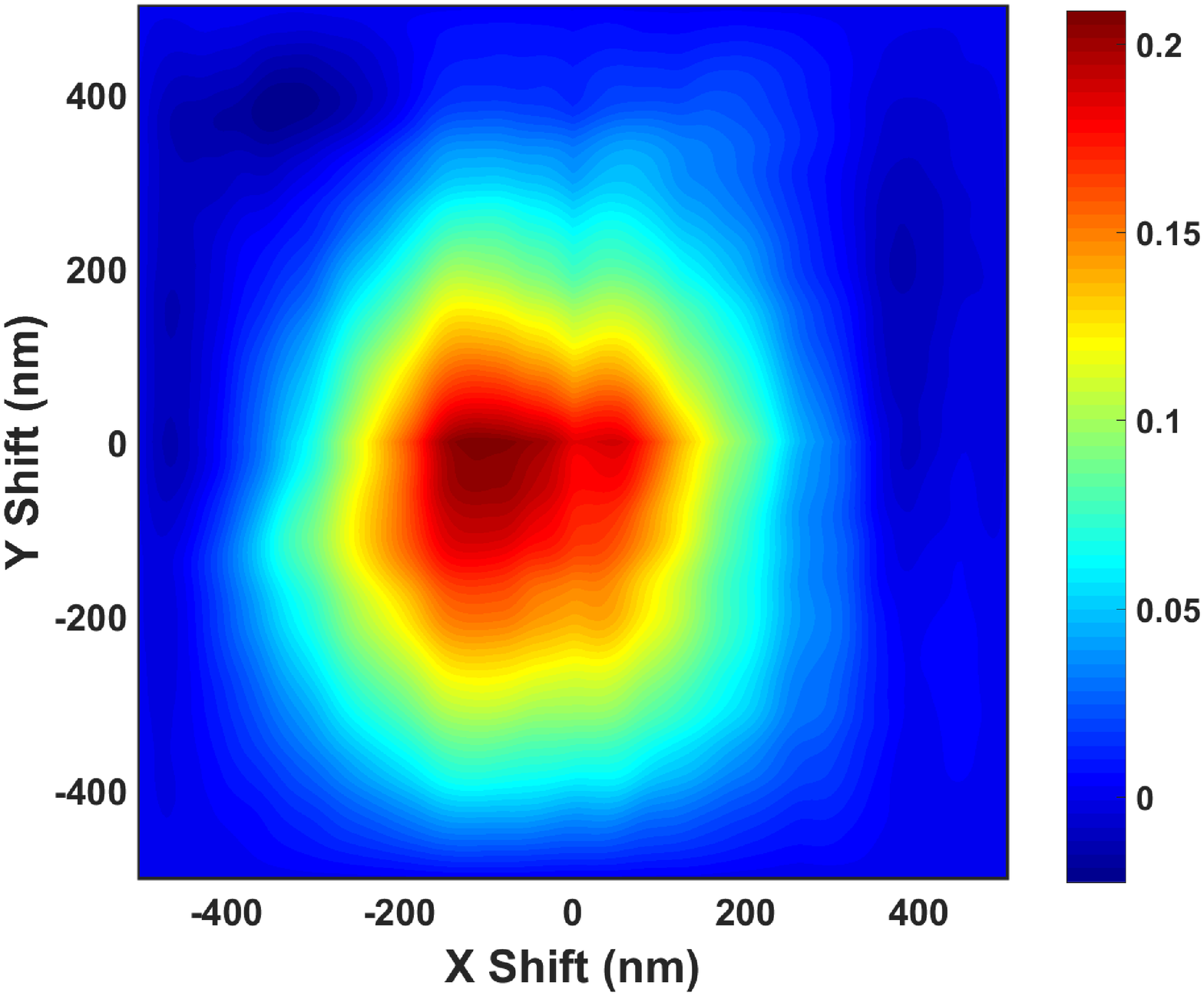}
        (f)
        \end{center}
    \end{minipage}
\caption{The topographies acquired using CITS with colormaps
according to the ZBC values for (a) 5~nm, (c) 10~nm, and (e)
100~nm MoGe samples. Two dimensional cross correlation between
topography and ZBC for 5~nm (b), 10~nm (d), and 100~nm~(f) MoGe
thin films. The measurements were performed at
0.53~K}\label{fig:crosscor}
\end{figure}
\twocolumngrid

\begin{equation}
\centering
 \frac{\Delta_n}{\Delta_s}=\left(1+\frac{L_n}{2.1\xi_n}\right)^{-2},
 \label{eq:egDelta}
\end{equation}
where $L_n$ is the thickness of normal metal. Taking as a true
superconducting gap in bulk MoGe 1.13~meV~\cite{Tashiro} and
minigap from our experiments, we receive the ratio
$\Delta_n/\Delta_s$=~0.90. From Eq.~\eqref{eq:egDelta} we receive
the ratio $L_n/\xi_n$=~0.08. It is in a good agreement with our
experimental results $L_n/\xi_n$=5~nm/60~nm =~0.08. Thus, for
explanation of the superconducting gap in Au layer a proximity
effect is sufficient. We can also conclude that down to 10~nm of
MoGe films their superconducting properties are close to that of
the bulk.

On 5~nm MoGe/Au bilayer the minigap $\Delta_n$ is strongly
suppressed. Also coherence peaks are almost absent and heavy ZBC
present. Since $L_n/\xi_n$ is not changed in comparison  to the
100 and 10~nm cases another pair breaking mechanism must operate
to explain the observed spectra. At 5~nm MoGe thickness second
interface is present above the STM junction, specifically
superconductor - substrate interface. At this interface additional
pair breaking effect - spin-flip scattering can be operational
\cite{Belzig}) leading to a thin layer with suppressed
superconducting parameters. This layer can affect the MoGe film
via an inverse proximity effect and cause a strong smearing of the
conductance spectra even at such low temperatures.

The ZBC maps obtained via current image tunneling spectroscopy are
plotted in Fig~\ref{fig:crosscor} together with a topography of
the Au overalayer. The $z$-axis corresponds to the topography
while colormap is the values of the ZBC
(Fig.~\ref{fig:crosscor}(a,c,e)). Such combination allows us to
define the influence of the roughness of the surface on the
superconducting properties. The two-dimensional cross-correlation
between topography and ZBC was calculated for each sample
(Fig.~\ref{fig:crosscor}(b,d,f)). The relatively high central peak
for 10 and 100~nm samples with the correlation value 0.2 indicate
a good correlation between topography relief and ZBC. The hills of
topography correspond to locally thicker gold film. Then, at these
places the superconductivity would be slightly suppressed. In
opposite, the pit parts have a thinner gold cover, and the lower
values of ZBC is expected. It provide another evidence that in the
10 and 100~nm MoGe films the proximity effect is dominant.

On the contrary, the 5~nm MoGe sample shows the weakest
correlation with a smeared central peak and the correlation value
of only 0.08. It is despite the high RMS roughness which should
impose a strong proximity effect. All this corroborates another
pair-breaking mechanism, most probably the spin-flip scattering at
the interface which is expanded by the the inverse proximity
effect to the bilayer

\section{Conclusions}
Scanning tunneling microscopy and spectroscopy measured on 5~nm
gold overalyer on 100, 10 and 5~nm MoGe films reveal suppression
of superconducting energy gap and $T_c$ upon reduction of the
thickness of MoGe films. The two-dimensional cross-correlation
between topography and ZBC for the 10 and 100~nm MoGe samples
indicates a proximity effect in Au layer where slightly smaller
superconducting energy gap as compared to the bulk MoGe is
observed. The superconducting properties of MoGe remain unchanged
down to 10~nm thickness. In the STM LDOS spectra of the 5~nm MoGe
sample a high ZBC and low coherence peaks are observed. This
cannot be explained by a simple proximity effect and another
mechanism is functioning. It is probably an inverse proximity
effect from the interface between the MoGe film and the glass
substrate where the spin-flip scattering suppresses the
superconductivity.

\section{Acknowledgments}

This paper is dedicated to Professor Viktor Valentinovich Eremenko
on the occasion of his 85th birthday. Authors (A.F. and P.S) are
very thankful to Prof. Eremenko for decades of fruitful
collaboration. Viktor Valentinovich was and still is an excellent
mentor in the field of magnetism for the whole Kosice Low
Temperature Group.

This work was supported by projects APVV-0605-14, VEGA~1/0409/15
and U.S.~Steel~Ko\v{s}ice.

\bibliography{ltp_bibliography}

\end{document}